\documentclass[11pt]{article}

\usepackage{times}
\usepackage[version=3]{mhchem}
\usepackage{authblk}
\usepackage{hyperref}
\usepackage[capitalize,noabbrev]{cleveref}
\usepackage{booktabs}
\usepackage{longtable}
\usepackage{amssymb}
\usepackage{lineno}
\usepackage{setspace}
\usepackage[dvipsnames]{xcolor}
\usepackage{geometry}
\usepackage{pdflscape}

\usepackage[title]{appendix}
\usepackage{natbib}
\bibpunct[]{(}{)}{;}{a}{,}{,}

\usepackage[dvipdfmx]{graphicx}

\graphicspath{{Fig/}}

\title{\textbf{Chemically defining the building blocks of the Earth}}
\author[1]{Takashi Yoshizaki\thanks{Corresponding author. E-mail: \href{mailto:takashiy@tohoku.ac.jp}{takashiy@tohoku.ac.jp}}}
\author[2]{Richard D. Ash}
\author[3]{Tetsuya Yokoyama}
\author[2]{Marc D. Lipella\thanks{Deceased.}}
\author[1,2,4]{William F. McDonough}
\affil[1]{Department of Earth Science, Graduate School of Science, Tohoku University, Sendai, Miyagi 980-8578, Japan}
\affil[2]{Department of Geology, University of Maryland, College Park, MD 20742, USA}
\affil[3]{Department of Earth and Planetary Sciences, Tokyo Institute of Technology, Ookayama, Tokyo 152-8851, Japan}
\affil[4]{Research Center of Neutrino Sciences, Tohoku University, Sendai, Miyagi 980-8578, Japan}

\begin{document}
	
	
	\maketitle
	
	\section*{Abstract}
	\label{abstract}
	%
	
	Chondrites are undifferentiated sediments of material left over from the earliest solar system and are widely considered as representatives of the unprocessed building blocks of the terrestrial planets. The chondrites, along with processed igneous meteorites, have been divided into two broad categories based upon their isotopic signatures; these have been termed the CC and NC groups and have been interpreted as reflecting their distinctive birth places within the solar system. The isotopic distinctiveness of NC and CC meteorites document limited radial-mixing in the accretionary disk. The enstatite and ordinary chondrites are NC-type and likely represent samples from inner solar system (likely $ < $4 AU).  Measurement and modeling of ratios of refractory lithophile elements (RLE) in enstatite chondrites establish these meteorites as the closest starting materials for the bulk of the silicate Earth and the core. Comparing chondritic and terrestrial RLE ratios demonstrate that the Bulk Silicate Earth, not the core, host the Earth's inventory of Ti, Zr, Nb, and Ta, but not the full complement of V.\bigskip
	
	\noindent Keywords: chondrites, building blocks of Earth, early solar system, refractory lithophile elements, Nb/Ta
	
	\section*{Highlights}
	\begin{itemize}
		\item Relative abundance of refractory lithophile elements (RLE) show up to $ \sim $10\% variation among chondrite classes.
		\item Enstatite chondrites and the bulk silicate Earth have similar ratios of RLE.
		\item Building blocks of the Earth were chemically and isotopically similar, but not identical to, enstatite chondrites.
	\end{itemize}
	
	\section{Introduction}

	Compositional models of the Earth assume refractory lithophile elements (RLE) are in chondritic proportions (\textit{constant RLE ratio rule}) in the silicate fraction of the planet. For the Earth this means the Bulk Silicate Earth (BSE). There is, however, considerable discussion about the behavior of Nb and Ta during core-mantle differentiation, with some authors suggesting that Nb and Ta are exclusively lithophile \citep{,sun1989chemical,mcdonough1991partial,rudnick2000rutile}, whereas, others have suggested that Nb and less so Ta behaved as a chalcophile \citep[]{munker2017silicate} and/or siderophile \citep[]{,wade2001earth,munker2003evolution,corgne2008metal,cartier2014redox} element and has been partially sequestered into the core. These latter models have based their conclusions on an assumed "chondritic" ratio for Nb/Ta in the BSE.\bigskip

	Chondritic meteorites, undifferentiated solar system materials that are widely considered the building blocks of the terrestrial planets, have been divided into two broad categories based upon their isotopic signatures and interpreted as reflecting their distinctive birth places within the solar system \citep{warren2011stable}. The specific chondritic building blocks for the Earth are unknown, but the Earth and enstatite chondrites (EC) share many isotopic ratios, which sets them apart from the other groups of chondrites \citep{javoy2010chemical, boyet2018enstatite}.\bigskip 
	
	Recent analyses of chondrites can now be used to define better the values for \textit{constant RLE ratios}. Bulk data for the CI carbonaceous chondrites (CC), those with closest compositional match to the solar photosphere, have been reported \citep{barrat2012geochemistry,braukmuller2018chemical}. \citet{stracke2012refractory} reported a series of experiments to evaluate the mass of sample needed to obtain a representative bulk rock composition in Allende. \citet{barrat2014lithophile} reported bulk compositional data for EC. Focused studies on elements and their ratios in groups of chondrites have been reported (e.g., (REE) \citep{pourmand2012novel, dauphas2015thulium}, Y/Ho \citep{pack2007geo}, Zr/Hf \citep{patzer2010zirconium}, Nb/Ta \citep{munker2003evolution}).  Here we report major and trace element data on bulk rock, chondrules, and sulfides from EC, chondrules from an ordinary chondrite (OC), and refractory inclusions (Ca,Al-rich inclusions (CAIs) and amoeboid olivine aggregates (AOAs)) from CV carbonaceous chondrites. Using these and published data, we examine the compositional variability in some chondritic ratios, the behavior of elements in the protoplanetary disk and during core-mantle differentiation, and the chondritic reference frame for the BSE.\bigskip
	

	\section{Background and historical perspective}
	
	The cosmochemical classification of elements is based on the partitioning behavior during condensation into a silicate (lithophile), metal (siderophile) or sulfide (chalcophile) phase, with an assumed reference frame of half-mass condensation temperatures from a solar nebula at 10 Pa density condition \citep{,lodders2003solar}. The refractory elements (e.g., Zr, Re, Os, Ca, Al) are those with nebular half-mass condensation temperatures $ > $1360 K. Importantly, the abundant elements, Mg, Si, Fe and Ni (along with O), which make up the mass of terrestrial planets, have nebular half-mass condensation temperatures of 1360--1290 K. Moderately volatile (e.g., K, S, Zn) and volatile elements (e.g, H, C, N, often considered as ices or atmophile elements) have lower condensation temperatures. Abundances of the moderately volatile elements in chondritic meteorites can show enrichments (in EC) or depletions (in CC) relative to the refractory elements when normalized to CI abundances \citep[]{wasson1988compositions}.\bigskip
	
	There are some 36 refractory elements, and for the Earth most of them are generally considered to be lithophile (e.g., Be, Al, Ca, Ti, Sc, Sr, Y,  Zr, Nb, Ba, REE, Hf, Ta, Th and U), while V, Mo, and W are moderately siderophile/chalcophile, and Ru, Rh, Re, Os, Ir, and Pt are highly siderophile. The \textit{constant RLE ratio rule} asserts that the RLE are in chondritic proportions and not fractionated by core segregation processes. An important question to be addressed, particularly in the light of greatly improved analytical precision, is what are our working definitions of "constant" ratios for element pairs. Here we provide additional perspective.\bigskip
	
	The REE are the archetypal example of the rule. The flat, unfractionated Masuda-Coryell plot of the REE \citep[chondrite-normalized plot of the REE; e.g.,][]{,masuda1957simple} is a poignant visual of the rule of the \textit{constant RLE ratio rule}, demonstrating that the relative abundances of the REE are invariant between chondrite classes and that this principle should be reflected in the BSE. The limited Sm/Nd variation ($<$1\%) and even narrower variation in \ce{^{143}Nd/^{144}Nd} values in chondrites \citep[e.g.,][]{,bouvier2008lu} reinforce the rule.\bigskip
	
	In the 1960s, \citet{,gast1960limitations} and \citet{,wasserburg1964relative} established that the Earth did not have chondritic proportions of the alkali metals, which are moderately volatile elements, and did so with respect to the RLE. In the 1970s and 1980s, with the advent of Nd \citep{,depaolo1976inferences} and Hf \citep{,patchett1980hafnium} isotope systems, we further refined the \textit{constant RLE ratio rule} demonstrating that the BSE has chondritic Nd and Hf isotopic compositions and also chondritic Lu/Hf. Using correlations of Hf and Nd isotope systems with \ce{^{87}Sr}/\ce{^{86}Sr} for mantle derived rocks (noting that Sr is a RLE) provided a constraint on the present day \ce{^{87}Sr}/\ce{^{86}Sr} isotopic ratio of the BSE (e.g., 0.704--0.706) and from that its Rb/Sr value (0.032 $\pm$ 0.007, with uncertainty representing the spread in Sr isotope ratios) \citep{McDonough92}.  \bigskip
	
	The concept of "constant" RLE ratios came into question recently (2005--2015), with the measurement of variable \ce{^{142}Nd}/\ce{^{144}Nd} in chondrites, which emboldened some to assert that the BSE did not possess chondritic proportions of Sm/Nd \citep[e.g.,][]{,boyet2005142nd}. Solutions to this conundrum included: (1) a sequestered, isolated reservoir in the silicate Earth that was created in the first few hundreds of million years, which re-balanced the Earth's Sm/Nd back to "chondritic", (2) the Earth experienced mass loss of differentiated early Hadean crust by collisional erosion processes, and/or (3) fractionation of Sm/Nd between the core and primitive mantle \citep[][ and references therein]{,caro2011early}. However, by 2016 these debates were silenced by the observations of coupled (\ce{^{142}Nd}, \ce{^{145}Nd},  \ce{^{148}Nd}, \ce{^{150}Nd} and \ce{^{144}Sm}) nucleosynthetic isotopic anomalies between groups of chondrites, implying preserved isotopic heterogeneities in the protoplanetary accretion disk \citep[e.g.,][]{,bouvier2016primitive,burkhardt2016nucleosynthetic,fukai2017neodymium}. 	Consequently, the rule remains intact, albeit with the recognition that there are isotopic differences between the chondrites. Thus, the \textit{constant RLE ratio rule} has limits on the level of what is meant as constant and, for the ratio of the REE, this concept of being constant does not extend down to the ppm level of precision.\bigskip
	
	Concurrently, \citet{,warren2011stable} defined two isotopically distinct compositional meteoritic domains (carbonaceous (CC) and non-carbonaceous (NC)) for chondritic and non-chondritic meteorites and in doing so established the \textit{Warren Gap}. To date there are more than 10 isotopic systems (e.g., O, Ca, Ti, Cr, Ni, Mo, Ru, Nd, W, Os) that reveal measurable differences between CC and NC groups, with some isotope systems showing gross compositional distinctions. The dynamical implications for radial transport across the protoplanetary accretion disk is significant \citep[e.g.,][]{,kruijer2017age}. Importantly, the Urey-Craig diagram, established in the 1950s \citep{,urey1953composition}, documents a redox gradient in meteorite groups and further elucidates that the \textit{Warren Gap} shows the CC-group are definitively more oxidized than the NC-group (\cref{fig:UC_diagram}).\bigskip
	
	Understanding the partitioning behavior of elements and their ultimate residence in the silicate shell or metallic core requires knowledge of the redox conditions in the protoplanetary accretion disk and/or during planetary core formation. The integration of these two environments determines whether an element is wholly or partially lithophile, with the latter state indicating that an element's chemistry has been partially dictated by the extraction of a siderophile (Fe-affinity) and/or chalcophile (S-affinity) phase into a planetary core.\bigskip

	\section{Samples and methods}
	
	Primitive EC, OC and CC from Antarctica were studied. Analyses were conducted on bulk samples, chondrules, and sulfides from Alan Hills (ALH) A77295 (EH3), ALH 84170 (EH3), ALH 84206 (EH3), ALH 85119 (EL3) and MacAlpine Hills 88136 (EL3) and chondrules from Queen Alexandra Range 97008 (L3.05). We also studied two CAIs (3529-63-RR1 and 3529-61-RR1) from Allende (CV3) and a CAI (R7C-01) and 6 AOAs from Roberts Massiff 04143 (CV3). All samples were chosen for the lack of or limited terrestrial weathering. To obtain representative bulk chondrite composition, large sample aliquots ($\sim$1 g) were used in this study. Extensive details of the petrography of the studied samples are reported as supplementary information.\bigskip
	
	Petrology, mineralogy and major element chemistry of polished sections of samples were analyzed using scanning electron microscopes (SEM) and electron microprobes at Tohoku University and University of Maryland, College Park (UMd). Trace RLE compositions of chondrules from EC (mostly porphyritic pyroxene chondrules; Tables E.2 and E.3) and OC, sulfides from EC (e.g., troilite, oldhamite; Tables E.5 and E.6), and  refractory inclusions from CC (Table E.7) were determined using a New Wave frequency-quintupled Nd:YAG laser system coupled to the Element2 ICP-MS at UMd. National Bureau of Standards (NIST) 610 glass (for EC chondrules and CC refractory inclusions), United States Geological Survey (USGS) reference glass BCR-2g and NIST 612 glass (for OC chondrules) were used as external standard materials. Internal standards were \ce{^{29}Si} for chondrules and \ce{^{29}Si}, \ce{^{43}Ca} or \ce{^{47}Ti} for refractory inclusions, which were determined as a polymineralic bulk composition using EPMA.	For sulfides, the group IIA iron meteorite Filomena (for siderophile elements) and the NIST 610 glass (for lithophile elements) were used as reference materials.	Count rates were normalized using \ce{^{57}Fe} (for Filomena) and \ce{^{63}Cu} (for NIST 610) as internal standards, which were determined using electron microprobe or energy-dispersive spectrometers equipped with the SEM.\bigskip
	
	Details of sample preparation and analytical technique are also provided in supplementary files. In short, about 25 mg of powder was weighed in Teflon vessels from $ \sim $1 g of homogenized powder. Two aliquots of each sample were prepared. The powder was mixed with \ce{^{91}Zr}-\ce{^{179}Hf}  and \ce{^{113}In}-\ce{^{203}Tl} spikes for isotope dilution-internal standardization mass spectrometry of HFSE and REE \citep{,lu2007coprecipitation,yokoyama2017investigating}, and Al solution to avoid precipitation of HFSE as fluoride compounds \citep{,tanaka2003suppression}. The sample was then digested by a combination of \ce{HCl}, \ce{HNO3}, \ce{HF} and a solution for HFSE measurement was isolated. The mother solution was then attacked with \ce{HClO4} and \ce{HNO3} and a split for REE analysis was separated. The HFSE and REE solutions were further divided into two aliquots for measurements at University of Maryland (UMd) (using Thermo Finnigan Element2 single-collector, sector field inductively coupled plasma mass spectrometer (ICP-MS)) and Tokyo Institute of Technology (Thermo Scientific X Series Q-pole ICP-MS) (Table E.1). A standard basaltic rock JB-3 (Geological Survey of Japan) was also digested and used as a reference material.\bigskip
	
	\bigskip
	
	\section{RLE in chondrites and their components}
\begin{figure}[p]
		\centering
		\includegraphics[width=1.0\linewidth]{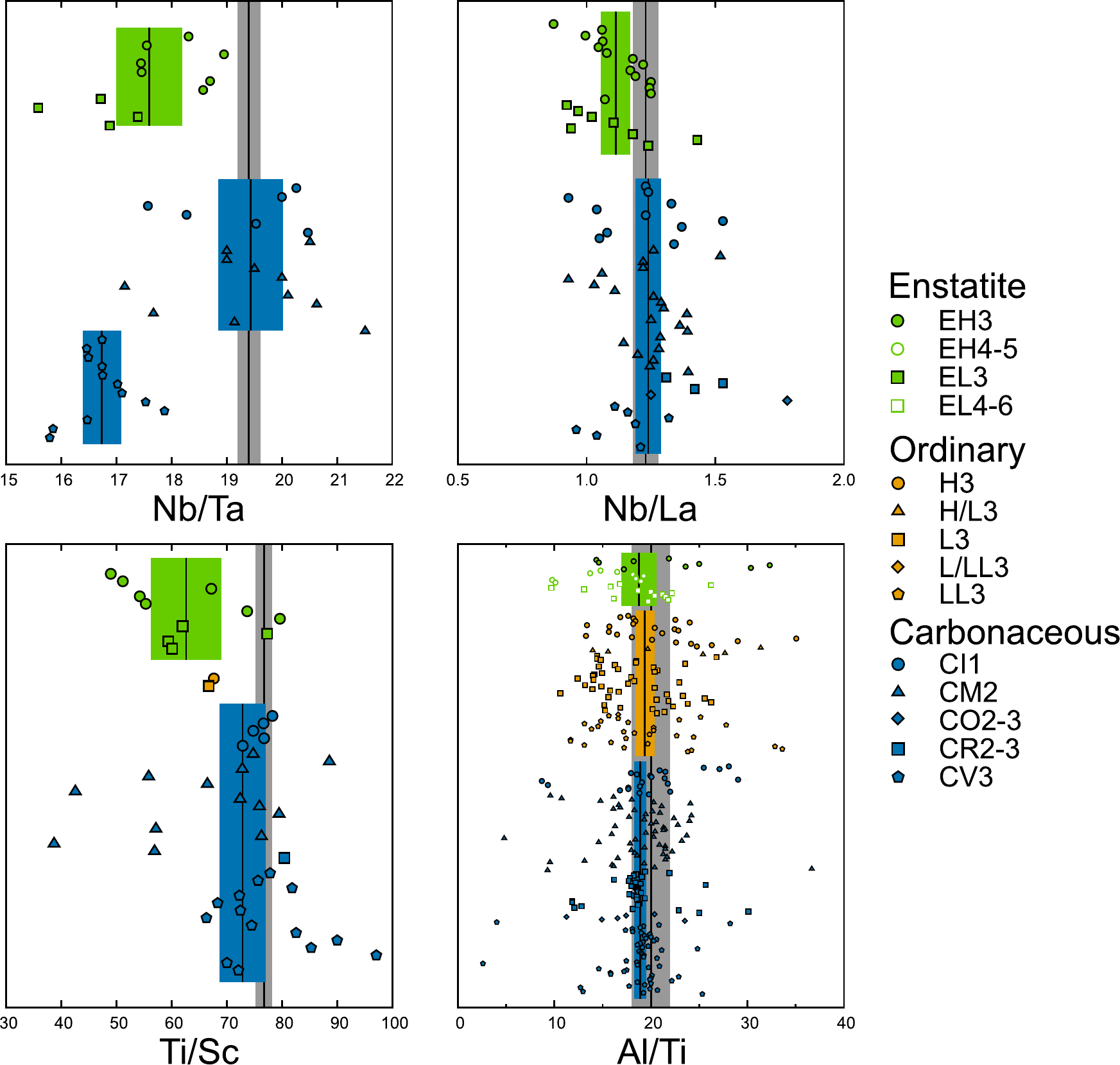}
		\caption{Refractory lithophile element ratios of unmetamorphosed, bulk enstatite (green), ordinary (orange) and carbonaceous (blue) chondrites. Colored box and inside solid line represent mean $ \pm ~ 2\sigma _{\mathrm{m}} $ values for each chondrite groups. Gray box and inside solid line correspond mean $ \pm ~ 2\sigma _{\mathrm{m}} $ for the Orgueil CI chondrite \citep{barrat2012geochemistry}. For Al/Ti ratio, data for metamorphosed enstatite chondrites are also shown and used to calculate the mean $ \pm ~ 2\sigma _{\mathrm{m}} $ value, since the number of available data for type 3 enstatite chondrites are limited. Sources of literature data are available as supplementary information.}
		\label{fig:chondrite_RLEratio}
	\end{figure}

	Most of the RLE are transition elements and some of these can have non-lithophile behavior in some systems (e.g., V). Here, we report on the relative abundances of the triad Sc-Ti-V and the higher Z element members of each {IUPAC} group in the Periodic Table (i.e., Y, REE  \textendash  ~Zr, Hf  \textendash ~Nb,Ta). Each of these element pairs are often considered as rarely fractionated, geochemical twins (i.e., Y/Ho, Zr/Hf and Nb/Ta). Bulk compositional data for primitive EC are presented, along with a literature compilation for bulk unequilibrated chondrites showing the mean values for several RLE ratios (e.g., Ti/Sc, Nb/Ta and Nb/La) (\cref{fig:chondrite_RLEratio,tab:RLE_chondrite_BSE}). Ratios of Nb/Ta and Ti/Sc, and less so for Nb/La, distinguish EC from CC, thus these ratios violate the \textit{constant RLE ratio rule}, or at least define a limit to the value of a ratio being constant.\bigskip

	Differences in refractory transition element ratios in different chondrite groups could reflect (1) volatility-driven fractionation of these elements in the protoplanetary disk and/or (2) non-lithophile behavior of one of these elements in the nebular disk or parent body in which metal-silicate fractionation occurred \citep[e.g.,][]{,kuebler1999sizes,wurm2013photophoretic}. What remains as an unknown is whether an element that behaved as non-lithophile in the disk keeps (or not) its non-lithophile character in the differentiating planet. Previously-reported variation in ratios of geochemical twins Y/Ho and Zr/Hf among chondrite groups \citep{,pack2007geo,patzer2010zirconium} become less clear but still observable when considering only unmetamorphosed samples (\cref{tab:RLE_chondrite_BSE} and Figure S.1), indicating that modification of these ratios occurred in the nebular environment.\bigskip
	
\begin{landscape}
		\begin{figure}[p]
			\centering
			\includegraphics[width=1.0\linewidth]{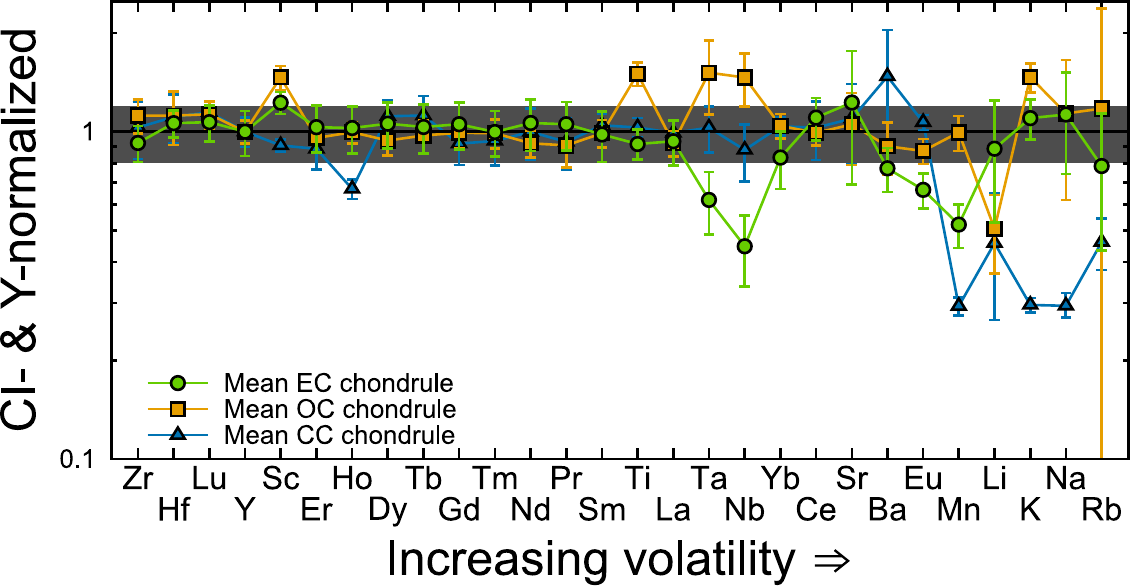}
			\caption{Mean abundance of lithophile elements in chondrules from each class of unequilibrated chondrite normalized to CI chondrite \citep{,barrat2012geochemistry} and Y. Elements are arranged in order of decreasing 50\% condensation temperature \citep{,lodders2003solar}. Error bars represent 2$ \sigma _{\mathrm{m}} $ and a gray box corresponds CI $ \pm ~20$\%. Sources of literature data are available as supplementary information.}
			\label{fig:chondrule_RLE_volatility}
		\end{figure}
	\end{landscape}

	A comparison of ratios of RLE in individual chondrules from EC, OC and CC provides evidence for compositional heterogeneity in the early solar system (\cref{fig:chondrule_RLEratio}). Although there is considerable compositional scatter in each group of chondrules, their means and standard errors provide a measure of distinction between classes of chondrites. There is no clear relationship between RLE ratios in chondrules and their petrological type. The Nb/Ta, Zr/Hf, Ti/Sc and possibly Sm/Nd ratios of EC chondrules appear to have the strongest offsets. Scatter in Sm/Nd for EC chondrules is larger than that in bulk rock samples (Figure S.1); low Sm/Nd values in chondrules are comparable with that seen in bulk rock data. EC chondrules show a weakly increasing trending in Sm/Nd with higher metamorphic grade. CI- and Y-normalized abundance of most RLE in chondrules ranges within CI $ \pm ~20\%$, whereas most non-refractory lithophile elements show depletion with a larger variation (\cref{fig:chondrule_RLE_volatility}).  When compared to CC chondrules, EC chondrules have lower Nb and Ta abundances and lower Nb/Ta values, whereas OC chondrules are enriched in Nb and Ta (\cref{fig:chondrule_RLE_volatility}).  \bigskip
	
	Refractory inclusions from CC show highly fractionated REE abundances, as well as for other RLE compositions (Figure S.3 and Table E.7), which reflect volatility-driven fractionation of RLE during condensation of these inclusions \citep[e.g.,][]{,boynton1975fractionation}. Ca,Al-rich inclusions (CAIs) show clearly lower Nb/Ta value (3--9) than all bulk CC and CAI-free AOAs, and most CC chondrules (Table E.7)).\bigskip
	
	Troilites from primitive EC are enriched in Ti, Nb and Zr (normally  lithophile elements) along with typical enrichment of chalcophile and siderophile elements (\cref{fig:FeS_RLE}). Thus, under highly reducing conditions, as observed in EC \citep[$\sim$IW$- 5$;][]{,wadhwa2008redox}, these nominally lithophile elements become chalcophile. On the other hand, Ta and Hf, which are generally considered to be geochemical twins of Nb and Zr, respectively, were not detected in troilites from EC, reflecting their strictly lithophile properties. There is no relationship between RLE abundance in these troilites and their 50\% condensation temperatures \citep{,lodders2003solar} as observed in the refractory inclusions \citep[e.g.,][]{,boynton1975fractionation}. Therefore, the chondrite-normalized abundances of Ti, Nb and Zr in these troilites reflects their relative chalcophilic behavior under reducing condition, making Ti more chalcophile-like than Nb and much more so than Zr. Additionally, no detectable amounts of RLE were found in Fe-Ni metal grains from an unequilibrated EH3 chondrite (Alan Hills A77295), indicating the weakly siderophilic behavior of these elements.  \bigskip
	
	\section{Discussion}
	
	\subsection{Nebular versus planetary fractionation of RLE ratios}
	
\begin{figure}[p]
		\centering
		\includegraphics[width=1.0\linewidth]{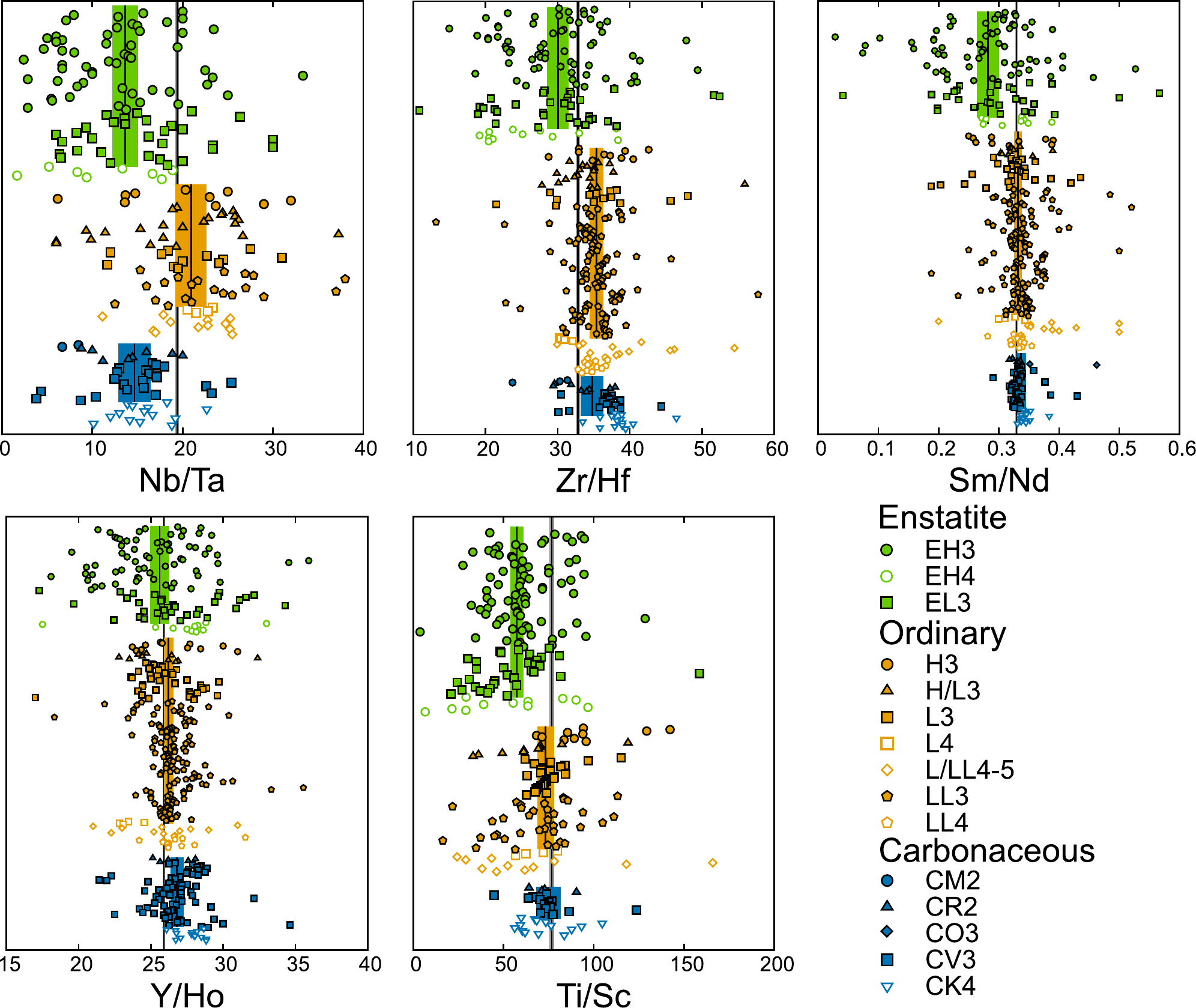}
		\caption{Refractory lithophile element ratios of bulk chondrules from each class of unequilibrated chondrite. Lines and boxes are as in \cref{fig:chondrite_RLEratio}. Data for chondrules from metamorphosed samples are also shown for a comparison. Sources of literature data are available as supplementary information.}
		\label{fig:chondrule_RLEratio}
	\end{figure}

	Chondrules, silicate droplets formed in the nebular disk, are one of the major RLE carriers in chondritic meteorites. Most chondrules are depleted in metals and from that it follows that metal/silicate fractionation occurred before/during chondrule formation \citep[e.g.,][]{,grossman1982mechanisms}. Moreover, it is well established that chondrule formation occurred locally (e.g., oxygen isotopic distinction between chondrules in different chondrite groups) with rare mixing of chondrules between distinct chondritic reservoirs before parent body formation \citep[e.g.,][]{,clayton2003oxygen_tog,hezel2018spatial}. Although there is considerable scatter, on average, CC and EC chondrules have lower Nb/Ta values and lower Nb and Ta abundances than bulk CI chondrites (\cref{fig:chondrule_RLEratio,fig:chondrule_RLE_volatility}). In contrast, chondrules from OC show mean Nb/Ta ratios higher than the CI value. EC chondrules have distinctly lower mean values for Zr/Hf, Ti/Sc and Sm/Nd as compared to OC and CC chondrules.\bigskip
	
	Since removal of Ti-bearing, Ta-free sulfide (\cref{fig:FeS_RLE}) from CI-like reservoir cannot produce the RLE pattern observed in EC chondrules, there should have been another mechanism that caused RLE fractionation in EC chondrules. We propose less refractory behavior of Nb and Ta under reducing conditions, with higher volatility of Nb than Ta. Although condensation chemistry of RLE at low $ f_{\ce{O2}} $ is still poorly constrained, \citet{,lodders1993lanthanide} showed that relative volatility of REE can be different under high and low C/O conditions. Further thermodynamic calculations are needed to understand fully changes of refractory behavior of RLE.\bigskip
	
	Compositional variations in chondrules from different chondrite groups reveal chemical fractionation prior to and/or during chondrule formation in the nebular disk. Combination of chemical (e.g., volatility, redox-dependent siderophility and chalcophility) and physical fractionation processes (e.g., sorting by density, size, magnetism, and/or thermal conductivity) of metal and silicates in the disk  
	\citep[e.g.,][]{,kuebler1999sizes,wurm2013photophoretic}  likely caused the observed variation in RLE ratios between chondrite groups. Chondrule populations, which likely accreted at different radial distances from the Sun, record the heterogeneous spatial distribution of chondrule source material with fractionated Nb/Ta ratio and/or localized fractionation of Nb and Ta. Thus, Nb/Ta variations among chondrite groups point to RLE fractionation before the accretion of chondrite parent bodies.\bigskip
	
	CV chondrites, recognized as having low bulk Nb/Ta values compared to other (mostly) CC \citep{,munker2003evolution,stracke2012refractory}, are unique among the CC in that they host abundant CAIs \citep[mean $ \sim $3 vol\%;][]{,hezel2008modal}, which formed as some of the earliest solids in solar system by high-temperature processes in a nebular gas. Small differences in condensation temperatures for Nb and Ta \citep[1559 K and 1573 K, respectively, at 10 Pa;][]{,lodders2003solar} could produce volatility-driven fractionation of these elements, resulting in the low Nb/Ta ratios in CAIs \citep[Table E.7;][]{,stracke2012refractory}. Thus, low Nb/Ta values in bulk CV chondrites document fractionation processes controlled by the nebular temperature.\bigskip
	
	Positive Tm/Tm* anomalies (deviations of measured Tm from expected Tm abundance derived from the interpolation between Er and Yb using a CI abundance), which is also caused by incorporation of CAIs, are detected in CC \citep[e.g.,][]{,dauphas2015thulium}. In contrast, the Earth, Mars, EC and OC, and several achondrites all share a slight ($\sim $3\%) negative Tm/Tm* anomalies \citep[e.g.,][]{,dauphas2015thulium}. Thus, volatility-dependent fractionation may be the origin for the observed variation in Nb/Ta and Nb/La ratios among the NC-group chondrites and the BSE.\bigskip
	
	There is $ \sim $10\% variation in Nb/RLE values in bulk chondrites, even without considering CV chondrites. These findings place limits on the use of the \textit{constant RLE ratio rule} when estimating planetary composition, but enables us to constrain the building materials of a planet based on its RLE composition. Among the three major chondrite classes, EC have the closest compositional match for Nb/RLE ratios to the terrestrial values, whereas those in CC are distinctly different from the Earth's, indicating a close genetic link between Earth and EC, as suggested by many isotope systems \citep[e.g.,][]{,dauphas2017isotopic}.   \bigskip
	
	Ratios of other geochemical twins, like Y/Ho and Zr/Hf, are also not constant between chondrites classes \citep{,pack2007geo,patzer2010zirconium}, which demonstrates the limits to the application of the \textit{constant RLE ratio rule}. It might be argued that there is no meteorite group which is chemically identical in RLE ratios as that of the Earth. Further investigations of RLE in meteoritic and terrestrial samples, including determination of Nb/Ta ratio of unequilibrated OC, can provide new insights into the origin of these heterogeneities in the protoplanetary disk. In turn these studies can be used as a new guide to constraining the composition of planetary building materials.  \bigskip
	
	These above studies begin to document that variations in ratios of RLE can be linked with isotopic systematics that establish the \textit{Warren Gap} and document compositional heterogeneities in the early solar system. Moreover, to understand the details of terrestrial geochemistry we endorse the idea \citep{dauphas2015thulium} that CC in general and CI chondrite specifically are not appropriate for normalizing refractory element abundances.\bigskip
	
\begin{figure}[p]
		\centering
		\includegraphics[width=0.9\linewidth]{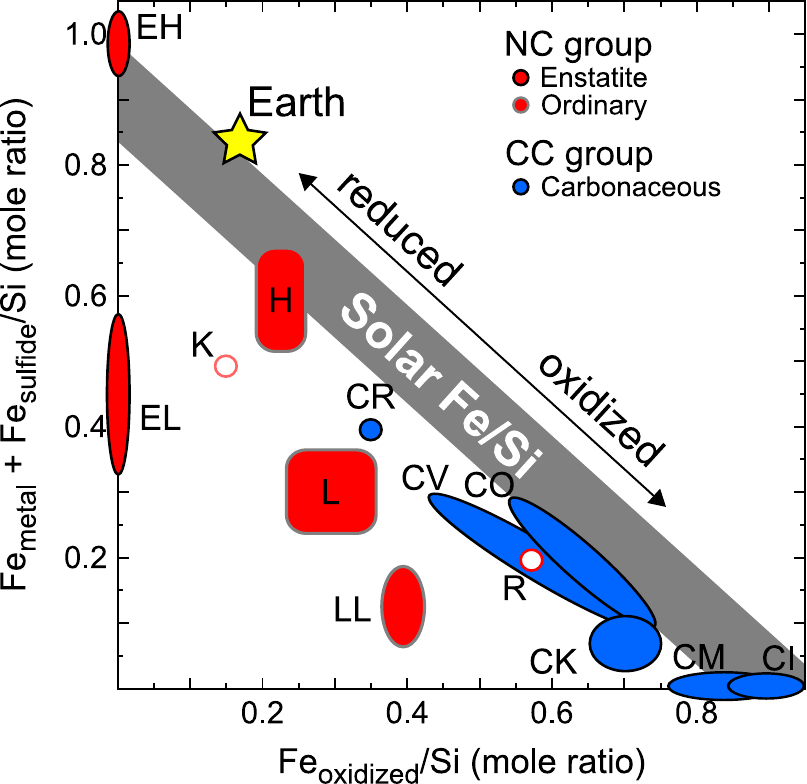}
		\caption{The Urey-Craig diagram showing variation in oxidation state of iron relative to silicon among chondrite groups \citep[after][]{,urey1953composition}. Metal-rich CB and CH chondrites are not included in the plot because their origin \citep[condensation from a gas-melt plume formed by a planetary-scale collision; e.g.,][]{,krot2012isotopically} is distinct from that of major chondrite groups. Non-carbonaceous (NC: enstatite and ordinary) and carbonaceous (CC) group chondrites are colored in red and blue, respectively, based on their isotopic compositions (\citealt{,warren2011stable} and references therein; \citealp{,budde2018early}). The Earth's composition is from \citet{,mcdonough2014compositional}. As recognized in many isotope systems, NC- and CC-groups are also distinct in their redox status except for Rumuruti (R)-type chondrites. Based on the redox condition of the Kakangari (K)-type chondrites, we predict that it is a member of the NC-group, which is supported by their low refractory element abundance and limited number of isotopic data \citep[][ and references therein]{,scott2018isotopic}.} 
		\label{fig:UC_diagram}
	\end{figure}

	Thus, the inner (e.g., Earth and NC-group meteorites) and outer solar system materials (CC-group meteorites) are likely to be distinctive, not only in isotopic composition \citep{,warren2011stable} and redox state (\cref{fig:UC_diagram}), but also in RLE composition. The higher Nb/Ta ratio of the bulk silicate Mars \citep[$ 20.1 \pm 0.9 $;][]{,munker2003evolution} indicates that Mars incorporated less EC-like material than Earth, which is consistent with multiple isotopic systematics \citep[e.g.,][]{,dauphas2017isotopic}. Primitive EC and the Moon have a similar Nb/Ta ratio \citep[$ 17.6 \pm 0.6 $ and $ 17.0 \pm 0.8 $, respectively: this study;][]{,munker2003evolution}, in harmony with the isotopic constraints that predict such an EC-like composition for the giant impactor that formed the Moon \citep[e.g.,][]{,dauphas2017isotopic}.  \bigskip
	
	Sulfide-silicate separation under highly reduced conditions, as in some regions of the nebular disk, may fractionate RLE in the early solar system. Sub-solar Fe/Si values seen in most chondrite groups, except for CI and EH chondrites (\cref{fig:UC_diagram}), document metal-silicate fractionation processes in the nebular disk. Likewise, it has been postulated that most chondritic asteroids preferentially sample the silicate-rich portion of nebular materials \citep[e.g.,][]{,wurm2013photophoretic}. Some amount of RLE fractionation (\cref{fig:chondrite_RLEratio,fig:chondrule_RLEratio}) may have been caused by separation of RLE-bearing sulfides from the solar-like reservoir in the nebular disk. Since only sulfides formed under highly reduced conditions can contain significant amounts of RLE \citep[e.g., this study;][]{,schrader2018background}, the bulk OC and CC composition is not affected by this process. Finally, given accretion of different chondrite groups took place in distinct time and/or space settings in the protoplanetary disk, we predict there to have been spatial and/or temporal variation in RLE ratios in the chondrite forming regions.\bigskip
	
	\subsection{Chemical composition of the Earth}
	
	Compositional modeling of the bulk Earth requires identifying which elements are hosted in BSE versus the core, which is done by comparing the chemical composition of the accessible parts of the Earth (crust and mantle) and undifferentiated chondritic meteorites. However, it is recognized that elements show different behaviors (siderophile, chalcophile, lithophile) in the solar nebula and during differentiation of a planet. Thus, combined geochemical and isotopic studies of chondritic and non-chondritic meteorites coupled with similar such studies of mantle samples and mantle-derived melts, are required to determine the composition of the BSE, which remains the best avenue for establishing the bulk Earth's composition.\bigskip
	
\begin{figure}[p]
		\centering
		\includegraphics[width=1.0\linewidth]{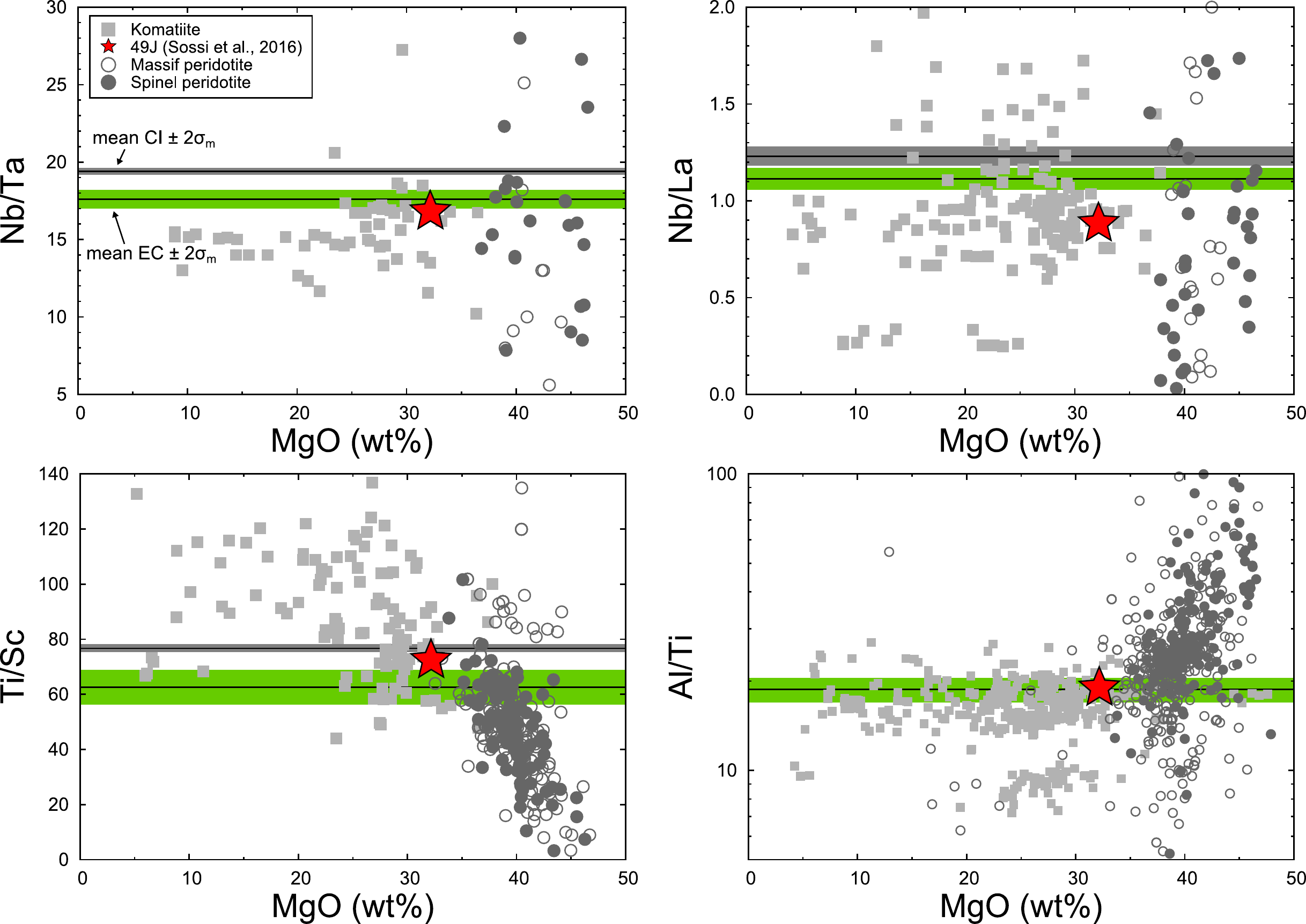}
		\caption{Comparison of refractory lithophile element ratio of the bulk silicate Earth (data from literature) and enstatite chondrites. Lines and boxes are as in \cref{fig:chondrite_RLEratio}. Red stars shows composition of the least altered komatiite 49J from Barberton, South Africa \citep{,sossi2016petrogenesis}, which might be close to the primitive mantle composition \citep[e.g.,][]{,sun1978petrogenesis,sossi2016petrogenesis}.}
		\label{fig:EC_vs_BSE}
	\end{figure}

	A comparison of chondritic ratios with that observed in primitive samples of the Earth's mantle and its melts is given in \cref{fig:EC_vs_BSE}. We highlight the composition of sample 49J, an Al-undepleted Barberton komatiite, which is a least altered, Paleoarchean mantle-derived melt having an exceptionally primitive composition \citep{,sossi2016petrogenesis} that samples a primitive mantle-like composition \citep[e.g.,][]{,sun1978petrogenesis}. The coincidence of Al/Ti, Ti/Sc, and Nb/Ta in 49J and the values for EC is notable. The offset between 49J and EC for Nb/La may reflect differences in the relative incompatibility of these elements, source depletion due to continental crust extraction, and/or differences between the Earth and EC. The data for komatiites and peridotites straddle the chondritic values for Ti/Sc and Al/Ti, reflecting melt-residue differentiation processes.  \bigskip
	
\begin{figure}[ht]
		\centering
		\includegraphics[width=1.0\linewidth]{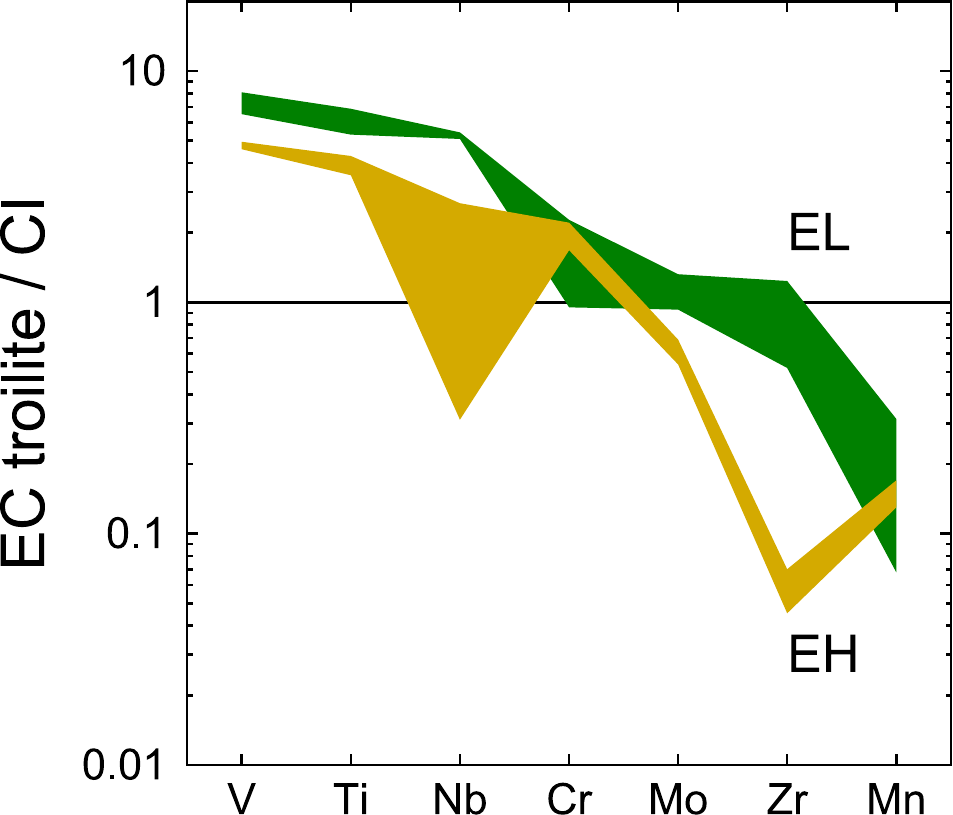}
		\caption{Ranges of mean transition element abundance in troilite from three EH3 (Alan Hills (ALH) A77295, ALH 84170 and ALH 84206) and two EL3 (ALH 85119 and MacAlpine Hills 88136) chondrites, normalized to CI chondrite composition \citep{,barrat2012geochemistry}.}
		\label{fig:FeS_RLE}
	\end{figure}

	It has been proposed that Nb and less so Ta have been depleted in the BSE due to their siderophile behavior under reducing conditions of core formation \citep[][]{wade2001earth,corgne2008metal,cartier2014redox}. Evidence against partitioning of Nb into the Earth's core due to its siderophile behavior comes from the composition of metals in meteorites and in experimental studies. In the reduced meteorites, Nb concentrations in metal (mostly $ \ll$0.1 ppm) are much lower than that in coexisting sulfides (typically $ > $0.4 ppm) \citep[this study;][]{,humayun2003microanalysis,kamber2003refined,van2012siderophile,munker2017silicate}, indicating that Nb is chalcophile rather than siderophile under reducing conditions. Partitioning experiments under low $ f_{\mathrm{\ce{O2}}}$ show that Ti, Nb, Ta and Ce can become more chalcophile than siderophile \citep{,wood2015trace,munker2017silicate}; similarly, these studies find that these elements are strongly partitioned into the sulfide relative to the co-existing metal phase. Therefore, if Nb is partitioned into a sulfide-bearing core \citep[1.5--2 wt\% S is in the Earth's core;][]{,mcdonough1995composition,dreibus1996cosmochemical} under reducing conditions, then a range of transition metals would have been sequestered into the sulfide portion of the core forming liquid. Finally, the common sulfide in reduced chondrites are Ti- and Cr-troilites \citep[this study;][ and references therein]{,van2012siderophile,weyrauch2017chemical}, and as we have found, these sulfides also contains Nb and lesser amounts of other transition elements (\cref{fig:FeS_RLE}).\bigskip
	
	\citet{cartier2014redox} observed that there is no pressure dependence on the siderophile behavior of Nb and Ta, which, if extrapolated to formation pressures of meteorites, would again argue for the chalcophile and not siderophile behavior for these elements. Moreover, a comparison of the experimental studies \citep[e.g.,][]{wade2001earth,corgne2008metal,cartier2014redox} shows that they have conflicting relative D-values for critical elements (e.g., Ni, Ga, Mn). This conflict is important as noted by \citet{corgne2008metal} who points out that it is possible to fit the data using an increasing $ f_{\mathrm{\ce{O2}}}$ during core separation, but that there is a delicate balance to get right the mass fraction, redox conditions and D-values to match the Earth's observed composition. There are diminishing probabilities of this highly determinsitic model when factoring in all of these findings.\bigskip
	
	In addition, the BSE has been proposed to be Nb-depleted due to its chalcophile behavior of Nb under reducing conditions \citep[][]{,munker2017silicate}, a conclusion based on the bulk Earth having a CC-like RLE composition. However, if Nb partitioning into the core was controlled by its chalcophile behavior, then combined Nb- and Ti-enrichments in EC troilites (\cref{fig:FeS_RLE}) document that Nb-depletion in the BSE should be accompanied by an even stronger Ti-depletion. In contrast, the coincidence of chondritic values for Ti/Sc and Al/Ti in the BSE and EC (\cref{fig:EC_vs_BSE}) indicate that the BSE hosts the Earth's inventory of both Nb and Ti, which is inconsistent with sequestering of Nb into the core via a sulfide.\bigskip
	
	The nature of an early Earth sulfide component and its migration to the core is poorly understood. Many proposals have been put forth for a wide range of elements (e.g., K, U, REE, RLE) partitioning into a sulfide phase and entering into the Earth's core during its formation \citep[e.g.,][]{,murthy2003experimental,munker2017silicate,wohlers2017uranium}. However, evidence is lacking for the role of a sulfide in controlling the inventory of these elements in the Earth 
	\citep{,corgne2007how,bouvier2016primitive,burkhardt2016nucleosynthetic,wipperfurth2018earth}.\bigskip
	
	The above models envisage extraction of Nb and perhaps Ta into the core based on a reference frame that assumes a CC model for the Earth's Nb/Ta value. A far simpler observation, and one consistent with  supporting stable and radiogenic isotopic data \citep{,warren2011stable}, is that the Earth has a Nb/Ta value comparable to that in EC. Thus, the BSE is not depleted in Nb compared to EC and the Earth's Nb inventory is not hosted in the metallic core. We conclude that the BSE and the bulk Earth have a shared Nb/Ta values of $ \sim $17.\bigskip

\begin{figure}[ht]
		\centering
		\includegraphics[width=1.0\linewidth]{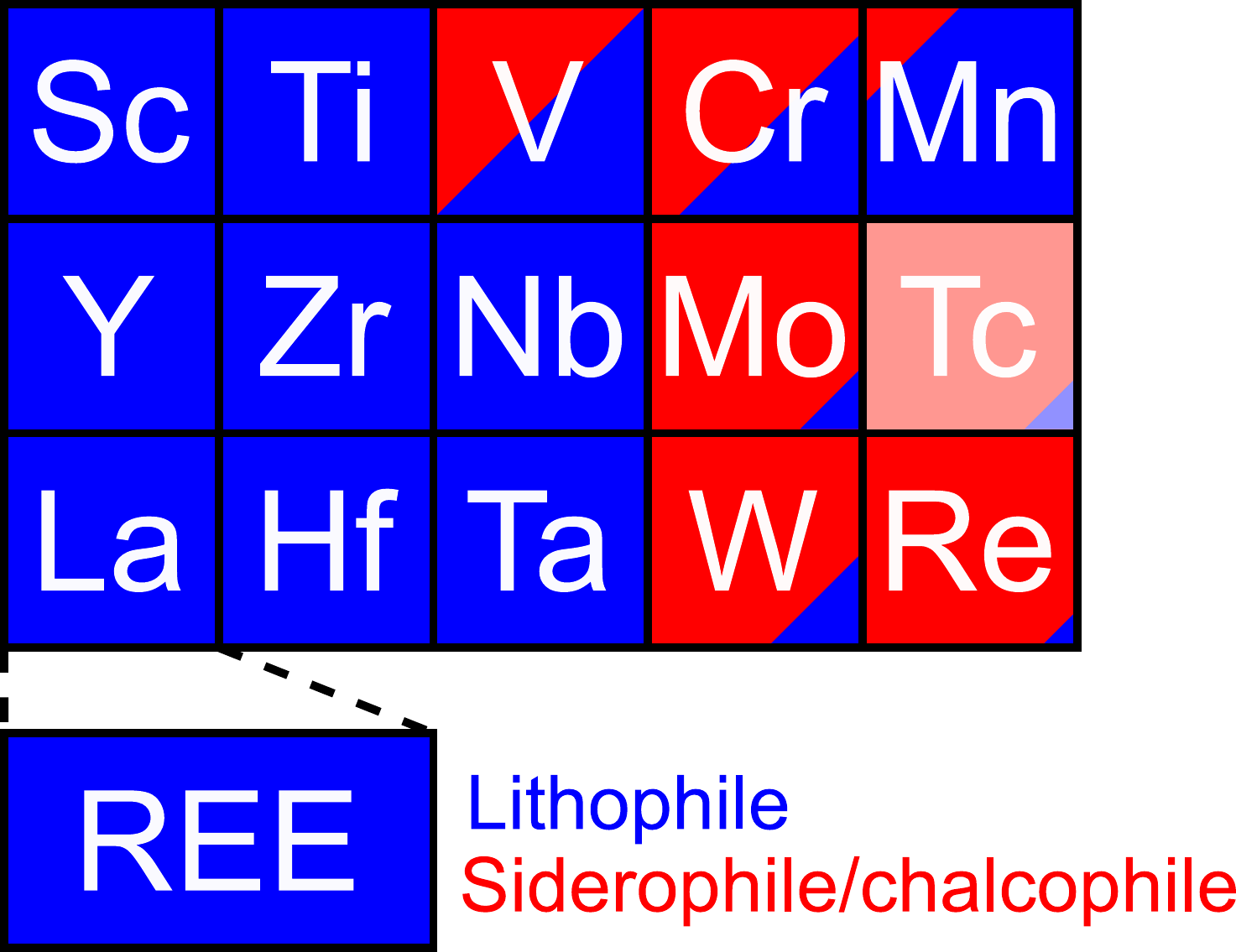}
		\caption{Geochemical classification of transition elements. Areas of blue and red correspond fraction of planetary inventory in the silicate Earth and the core, respectively. Data for V, Cr, Mn, Mo, W and Re are taken from Table 6 of \citet{,mcdonough2014compositional}. We expect that most fractions of Tc were partitioned into the core if its short-lived isotopes had been present in the planetary building materials.}
		\label{fig:RLE_periodic_table}
	\end{figure}

	
	Our findings allow us to characterize the chemical behavior of the left half of the transition metals in the periodic table (groups 3--7, using IUPAC nomenclature). Elements in each group ofttimes have the same electron configurations in their valence shells. However, during core-mantle differentiation these elements displayed markedly different geo-/cosmochemial (lithophile versus siderophile/chalcophile) behavior. The relative differences in their chemical behavior reflects their sensitivity to the oxidation state of their environment, which means changes in their electron configurations within a group. \cref{fig:RLE_periodic_table} is color coded in terms of mass fraction of the planet's budget of these elements in the BSE (blue) or core (red). In the Earth, 50\% of the planet's inventory of V is in the core and 50\% in the BSE \citep{,mcdonough2014compositional}. Elements in groups 3 and 4 and Nb and Ta (from group 5) all seem to have been lithophile during core-mantle differentiation and remained in the BSE. Also, the latter Period 4 (first row) transition metals have marked differences in their behavior (V, Cr, and Mn show variable affinities for the core), compared to those in Periods 5 and 6. For example, Mo is known to be more chalcophile than W, which is relatively more siderophile, and Re (and Tc?) is highly siderophile, meaning that is $ \sim $99\% of the planet's budget is in the core \citep{,mcdonough2014compositional}.     \bigskip

	Resolvable chemical and isotopic distinctions between chondrites (including EC) and Earth indicate that the building blocks of the Earth are not in our meteorite collection \citep[e.g.,][]{,mcdonough1995composition,drake2002determining}. Chondrites are residual solar system materials that originate from primitive asteroids that escaped any melting and differentiation processes. Their parent bodies formed some $\geq $2 Myr after solar formation \citep[e.g.,][]{hopp2016xe,blackburn2017accretion}, when radiogenic heating from \ce{^{26}Al} was waning. In addition, meteorites collected on the Earth are, except in rare instances, providing a restricted sampling of the asteroidal belt (2.0--3.5 AU) over the last $ < $100 Myr \citep{,heck2017rare}. Not only are these meteorites sampling a region of the nebula far from the Earth, they are likely further biased because of 4.5 billion years of harvesting of the asteroid belt by orbital resonances with Jupiter \citep{demeo2014solar}. Therefore, we are faced with considerable challenges when asserting the representativeness of materials present in the earliest stage of solar system formation. It is likely that the Earth accreted from inner solar system materials that were chemically and isotopically similar, but not identical to, EC, given chemical and isotopic heterogeneities and fractionation processes in the protoplanetary disk.\bigskip
	
	\section{Conclusions}
	
	We have found that Nb/Ta values differ between different groups of chondrites, likely due to redox conditions in the protoplanetary disk. We revealed that the BSE's Nb/Ta value ($\sim$17) matches that seen in enstatite chondrites ($\sim$17).  The Bulk Silicate Earth, not the core, host the Earth's inventory of Ti, Zr, Nb, and Ta, but not the full complement of V.\bigskip
	
	These findings provide insight into what are the limits of the \textit{"constant" RLE ratio rule}. We conclude that different ratios of RLE have greater variability than others (e.g., ratios involving only the REE show less variation than ratios involving Nb and/or Ta relative to other RLE). Thus, the chondritic record documents both chemical and isotopic variability in the nebular disk that can be used to inform dynamical model of stirring in the disk.  \bigskip
	
	Similar to the findings of \citet{munker2017silicate}, we find that sulfides in reduced meteorites host Nb, but importantly we also observed that these phases contain great amounts of Ti, which, if subtracted into the core, would have produced non-chondritic Al/Ti, Ti/Sc, and Ti/Zr values in the BSE and these are not observed. What remains as an unknown is whether an element that behaved as non-lithophile in the disk maintains (or not) its non-lithophile character during core separation.\bigskip
	
	Further studies of RLE ratios in chondrites are warranted, particularly those that focus on identifying chemical differences and similarities between groups of chondrites. A challenge which remains is understanding what are the causes for the differences in RLE ratios found between the Earth and chondrites. Are they due to nebular disk conditions, including phase separation, or to core-mantle differentiation processes?

	\section*{Acknowledgments}
	
	We thank R. Rudnick and P. Piccoli for discussions and technical assistance, I. Narita, S. Kagami, R. Fukai and I. Puchtel for technical assistance, and D.C. Hezel for kindly providing the ChondriteDB spreadsheet. We are grateful to NASA/JSC for loan of ALH 84170, ALH 84206, ALH 85119, MAC 88136 and RBT 04143, National Institute of Polar Research for ALHA77295, and the Smithsonian Institute for QUE 97008 and Allende. This work was supported by Grant-in-Aid for JSPS Research Fellow (No.~JP18J20708), GP-EES Research Grant and DIARE Research Grant to TaY, NSF grant EAR1650365 to WFM, and JSPS KAKENHI Grant (No.~16H04081) to TeY.
	
	\section*{Author contributions}
	WFM proposed and conceived the study with input from TaY. TaY conducted nearly all of sample preparation and measurements, with MDL performing some of the electron microprobe measurements, RDA and MDL conducting some of the LA-ICP-MS analyses and TeY performing some parts of sample digestion. Quantitative modeling was carried out by TaY in consultation with WFM. TeY, RDA, and WFM provided laboratory facilities and training to TaY. The manuscript was written by TaY and WFM, and all the authors contributed to it.
	
	\section*{Competing financial interests}
	
	The authors declare no competing financial interests.
	
	\clearpage
	
	\clearpage
	
	\clearpage
	
	\clearpage
	
	\clearpage
	
	\clearpage
	
	\clearpage
	
	\clearpage
	
	\begin{landscape}
		\begin{table}[p]
			\centering
			\footnotesize
			\caption{Refractory lithophile element ratios of bulk unequilibrated chondrites and the bulk silicate Earth (BSE). Sources of literature data are available as supplementary information.}
			\begin{tabular}{cccccccccccccccc}
				\toprule
				Class     &                                      &                      \multicolumn{2}{c}{Al/Ti}                       & Ti/Zr & Ti/Sc & Ti/V & Sc/V & Sc/Yb &           \multicolumn{2}{c}{Nb/Ta}           & Nb/La & Zr/Hf & Zr/Nb & Y/Ho & Sm/Nd \\
				\midrule
				Enstatite   &                 Mean                 & \multicolumn{1}{c}{21.9} & \multicolumn{1}{c}{18.8$ ^{\mathrm{a}} $} &  113  &  63   & 7.5  & 0.12 & 35.1  &           \multicolumn{2}{c}{17.6}            & 1.11  & 33.0  & 13.4  & 26.8 & 0.324 \\
				&             2$ \sigma $              & \multicolumn{1}{c}{13.0} & \multicolumn{1}{c}{10.4$ ^{\mathrm{a}} $} &  12   &  21   & 1.3  & 0.03 & 12.2  &            \multicolumn{2}{c}{2.0}            & 0.27  &  2.3  &  2.5  & 5.0  & 0.015 \\
				&      2$ \sigma _{\mathrm{m}} $       & \multicolumn{1}{c}{4.3}  & \multicolumn{1}{c}{1.8$ ^{\mathrm{a}} $}  &   4   &   6   & 0.4  & 0.00 &  1.8  &            \multicolumn{2}{c}{0.6}            & 0.06  &  0.5  &  0.5  & 1.1  & 0.003 \\
				&                $ n $                 &            9             &           33$ ^{\mathrm{a}} $            &  11   &  11   &  11  &  39  &  47   &            \multicolumn{2}{c}{11}             &  22   &  25   &  23   &  22  &  29   \\
				&                                      &   \multicolumn{1}{c}{}   &                                           &       &       &      &      &       &                       &                       &       &       &       &      &  \\
				Ordinary   &                 Mean                 &                       \multicolumn{2}{c}{19.4}                       &  94   &  67   & 8.5  & 0.12 & 36.4  &           \multicolumn{2}{c}{NA}                             &    NA   & 38.4  &   NA    & 26.1 & 0.332 \\
				&             2$ \sigma $              &                       \multicolumn{2}{c}{10.1}                       &  32   &   1   & 2.6  & 0.04 & 7.8  &                       &                       &       &  7.5  &       & 0.1  & 0.038 \\
				&      2$ \sigma _{\mathrm{m}} $       &                       \multicolumn{2}{c}{1.0}                        &  13   &   1   & 1.3  & 0.01 &  1.5  &                       &                       &       &  3.1  &       & 0.1  & 0.007 \\
				&                $ n $                 &                       \multicolumn{2}{c}{103}                        &   6   &   2   &  4   &  39  & 28   &                       &                       &       &   6   &       &  3   &  30   \\
				&                                      &   \multicolumn{1}{c}{}   &                                           &       &       &      &      &       &                       &                       &       &       &       &      &  \\
				Carbonaceous &                 Mean                 &                       \multicolumn{2}{c}{18.9}                       &  107  & 73   & 9.1  & 0.12 & 35.8  & 19.4$ ^{\mathrm{b}} $ & 16.8$ ^{\mathrm{c}} $ & 1.24  & 35.5  & 13.7  & 27.5 & 0.328 \\
				&             2$ \sigma $              &                       \multicolumn{2}{c}{8.6}                       &  42  & 24   & 2.7  & 0.03 &  9.1  & 2.4$ ^{\mathrm{b}} $  & 1.2$ ^{\mathrm{c}} $  & 0.34  & 11.5  &  3.2  & 5.9  & 0.038 \\
				&      2$ \sigma _{\mathrm{m}} $       &                       \multicolumn{2}{c}{0.7}                        &   5   &  4   & 0.3  & 0.00 &  0.8  & 0.6$ ^{\mathrm{b}} $  & 0.3$ ^{\mathrm{c}} $  & 0.05  &  1.2  &  0.4  & 0.8  & 0.003 \\
				&                $ n $                 &                       \multicolumn{2}{c}{172}                        &  71   &  33  &  82  & 121  &  145  & 17$ ^{\mathrm{b}} $  & 12$ ^{\mathrm{c}} $  &  47   &  95   &  80   &  56  &  146  \\
				\midrule
				\multicolumn{2}{c}{Komatiite 49J$ ^{\mathrm{d}} $ } &                       \multicolumn{2}{c}{18.9}                       &  109  &  72   & 11.1 & 0.15 & 32.3  &           \multicolumn{2}{c}{16.8}            & 0.88  & 38.3  & 16.5  & 28.1 & 0.326 \\
				&                                      &   \multicolumn{1}{c}{}   &                                           &       &       &      &      &       &                       &                       &       &       &       &      &  \\
				\multicolumn{2}{c}{BSE$ ^{\mathrm{e}} $ }      &                       \multicolumn{2}{c}{19.5}                       &  115  &  74   & 14.7 & 0.20 & 36.8  &           \multicolumn{2}{c}{17.8}            & 1.01  & 37.1  & 16.0  & 28.9 & 0.325 \\
				\bottomrule
			\end{tabular}	\\
			\flushleft
			NA--not available.\\
			a) includes metamorphosed samples \citep{,mason1966enstatite,von1968composition,wiik1969regular,jarosewich1990chemical,yanai1995catalog}.\\
			b) except for CV chondrites \citep{,jochum2000niobium,munker2003evolution,barrat2012geochemistry,gopel2015mn,friend2017composition}.\\
			c) CV chondrites \citep{,jochum2000niobium,munker2003evolution,lu2007coprecipitation,barrat2012geochemistry,stracke2012refractory}.\\
			d) from \citet{,sossi2016petrogenesis}. \\
			e) from \citet{,mcdonough1995composition}. \\
			\label{tab:RLE_chondrite_BSE}%
		\end{table}%
	\end{landscape}

	\begin{appendices}
		
		\section{Supplementary materials}
		Supplementary materials associated with this article can be found in 		the online version.
	\end{appendices}

	\bibliographystyle{elsarticle-harv}
	\bibliography{myrefs}

\end{document}